\documentclass[11pt]{article}
\usepackage{amsfonts}

\usepackage{amscd}
\usepackage{amsmath}
\usepackage{amscd}
\usepackage{amsmath}
\newlength{\dinwidth}
\newlength{\dinmargin}
\newtheorem{remark}{Remark}

\newtheorem{theorem}{Theorem}

\newtheorem{lemma}{Lemma}
\newtheorem{proposition}{Proposition}

\def\L{{\mathcal L}}
\def\M{{\mathcal M}}

\def\K{{\mathcal K}}

\def\res{\mathop{\rm res}\limits}

\begin{document}
\title{On the isomonodromic tau-function for the Hurwitz spaces of branched coverings of genus zero and one}
\author{Alexey Kokotov\footnote{e-mail: alexey@mathstat.concordia.ca},
Ian A. B. Strachan\footnote{e-mail: i.strachan@maths.gla.ac.uk}
}

\date{\today}
\maketitle
\centerline{$^\star$ Department of Mathematics and Statistics, Concordia University,}
\centerline{7141 Sherbrooke West, Montreal H4B 1R6, Quebec, Canada}

\bigskip

\centerline{$^\dagger$ Department of Mathematics, University of Glasgow}
\centerline{Glasgow, G12 8QW, United Kingdom.}

\bigskip

\begin{abstract}
The isomonodromic tau-function for the Hurwitz spaces of branched coverings
of genus zero and one are constructed explicitly. Such spaces may be equipped
with the structure of a Frobenius manifold and this introduces a flat
coordinate system on the manifold. The isomonodromic tau-function, and in
particular the associated $G$-function, are rewritten in these coordinates
and an interpretation in terms of the caustics (where the multiplication is
not semisimple) is given.
\end{abstract}
\def\Eta{{\bf \Upsilon}}
\def\B{{\mathbb B}}
\def\L{{\cal L}}
\def\l{\lambda}
\def\P{{\wp}}
\def\E{\tilde{\eta}}


\bigskip

\section{Introduction}

The equations which define the monodromy preserving deformations
of the Fuchsian system of differential equations
\begin{equation}
\frac{d\Psi}{d\l}+\sum_{k=1}^M\frac{A_k(\l_1, \dots, \l_M)}{\l-\l_k}\Psi=0
\label{Schlesinger}
\end{equation}
were derived by Schlesinger in 1912 and since then, particularly
after the seminal work of Jimbo and Miwa, they have been the
source of inspiration for many mathematicians; the equations
having many beautiful analytic, algebraic and geometric
properties. Naturally associated to such a system is a closed
1-form $\omega$ from which one may define a function $\tau$ via
$\omega=d\log\tau\,.$ It is this function that is the central
object of study in this paper.

A recent application of Schlesinger's equations has come from the
theory of Frobenius manifolds. These were introduced by Dubrovin \cite{Dubrovin}
as a geometric way to describe the associativity equations that arise in
topological quantum field theory (the
so-called Witten-Dijkgraaf-Verlinde-Verlinde equations). For semisimple
Frobenius manifolds one may reformulate the manifold's defining
relations as a reduction of Schlesinger's equations. In this
specific context these reduced equations have two interesting properties:

\begin{itemize}
\item{} the appearance of \lq wild\rq~monodromies and
Stokes matrices;
\item{} the existence of an alternative, distinguished,
coordinate system - the so-called flat coordinates $\{t^i\}\,.$
\end{itemize}
The associated $\tau$-function for such systems will be written
as $\tau_I\,$ and called the isomonodromic $\tau$ function.
This function also plays a major role in the
formula, conjectured by Givental \cite{G} and proved by
Dubrovin and Zhang \cite{DZ} for the $G$-function of
a semisimple Frobenius manifold
\begin{equation}
G=\log\left\{ \frac{\tau_I}{J^\frac{1}{24}} \right\}
\label{Gfunction}
\end{equation}
where
\begin{equation}
J=\det\left( \frac{\partial(t^1\,,\ldots\,,t^n)}{\partial(\lambda^1\,,\ldots\,,\lambda^n)}\right)
\label{Jacobian}
\end{equation}
is the Jacobian of the tranformation between the flat coordinates $\{t^i\}$ and
the canonical coordinates $\{\lambda^i\}\,.$ This function plays a number of
roles in the theory of Frobenius manifolds all related with the
construction of first order (or genus 1) objects from zeroth order (or genus 0)
data. For example, it appears in the genus 1-term in the free energy of the
related topological quantum field theory, in the generating functions for
genus 1-Gromov-Witten invariants, in first order terms in the expansion of oscillatory integrals
and in first order deformations of biHamiltonian structures.

A particular class of Frobenius manifolds may be constructed on Hurwitz
spaces - moduli spaces of meromorphic functions on Riemann surfaces.
(see \cite{Dubrovin} and section 2).
In this paper the $\tau_I$ and $G$-functions for genus 0 and 1 Hurwitz spaces
are studied. In section 2 these functions will be constructed in terms of critical
data on the Riemann surface. In section 3 the formulae will be re-evaluated
in terms of flat-coordinates, and this enables certain global properties of
these functions to be studied.

It will turn out that the non-semisimple part of the manifold plays a pivotal
role. By definition, a massive Frobenius manifold $M$ has a semisimple multiplication
on the tangent space at {\sl generic} points of $M\,.$ The set of points where the multiplication
is not semisimple is known as the caustic, and will be denoted $\K\,.$ This is an
analytic hypersurface in $M\,,$ which may consist of a number of
components (possibly highly singular),
\[
\K = \bigcup_{i=1}^{\#\K_i} \K_i\,.
\]
This forms part of the bifurcation diagram $\mathcal{B}=\{ {\mathbf\lambda}\,:\lambda^i-\lambda^j=0\,,i\neq j \}$
and in general $\mathcal{B}\cong\mathcal{K}\cup\mathcal{M}$ where $\mathcal{M}$ is the so-called
Maxwell strata.
The components of the caustic are given in terms of quasi-homogeneous irreducible polynomials $\kappa_i$
such that $\kappa_i^{-1}(0)=\K_i\,.$ For examples, using the ideas developed
in \cite{Strachan} one may show that the isomonodromic $\tau$-function
for the Frobenius manifold $\mathbb{C}^N/W$ where $W$ is a Coxeter group,
is given by
$$
\tau_I= \prod_{i=1}^{\#\K_i} \kappa_i^{-\frac{(N_i-2)^2}{16 N_i}}
$$
where the $N_i$ are certain integers determining the F-manifold structure
on the caustics \cite{H1}. Such a formula determines certain global
properties of the function, such as its zeros and poles and its irreducibility
properties. The case $W=A_{N-1}$ (where $\#\K_i=1$ and $N_1=3)$
coincides with the particular Hurwitz space $H_{0,N}(N)\,.$

\section{The isomonodromic tau-function}

In this section we give a short proof of the formulae for the isomonodromic tau-function
of semisimple Frobenius manifolds associated to the Hurwitz spaces of branched coverings of genus $0$ and $1$.
Closely related formulae for the isomonodromic tau-function of the class of
Riemann-Hilbert problems solved in \cite{Korotkin}
were first found in \cite{KokKor} without reference to any associated Frobenius manifold. The connections
between the formulae in \cite{KokKor} and Frobenius manifolds were conjectured in
\cite{Strachan} and finally established in \cite{KokKor1}.
These formulae were proved indirectly; the method was based on a detailed study of
the behaviour of some
regularized Dirichlet integrals under deformations of the branched coverings.
The introduction of Dirichlet integrals was inspired by the paper
\cite{ZT} on accessory parameters which appeared, at least superficially, to deal with a
similar question.

\medskip

Such methods use auxiliary structures not directly connected with the
holomorphic geometry of Hurwitz spaces.
It was conjectured in \cite{KokKor} that such indirect methods are inappropriate
and can be replaced by a direct one which should use only holomorphic geometry. Here we present such a proof.

\subsection{Preliminaries}

\subsubsection{Hurwitz spaces and Frobenius manifolds}\label{sec1}
Frobenius manifolds related to Hurwitz spaces were introduced by Dubrovin
(see \cite{Dubrovin}, chapter 5). It will be assumed that the reader is familiar
with this construction. Here we recall some basic facts and definitions in order to fix notation.

Let $H_{g, N}(k_1, \dots, k_l)$ be the Hurwitz space\footnote{
Dubrovin uses a slightly different notation. In his notation the Hurwitz space is
$H_{g;k_1-1\,,\ldots\,,k_{l}-1}\,.$}
of equivalence classes $[p:\L\rightarrow {\mathbb P}^1]$ of $N$-fold branched coverings $p:\L\rightarrow {\mathbb P}^1$, where
$\L$ is a compact Riemann surface of genus $g$ and the holomorphic map $p$ of degree $N$ is subject to the following conditions:
\begin{itemize}
\item it has $M$ simple ramification points $P_1, \dots, P_M\in\L$ with distinct {\it finite}
images $\l_1, \dots, \l_M\in {\mathbb C}\subset {\mathbb P}^1$;
\item the preimage $p^{-1}(\infty)$ consists of $l$ points: $p^{-1}(\infty)=\{\infty_1,
\dots,\infty_l\}$, and the ramification index of the map $p$ at the point $\infty_j$ is
$k_j$ ($1\leq k_j\leq N$).
\end{itemize}

\noindent (We define the ramification index at a point as the number of sheets of
the covering which are glued together at this point. A point $\infty_j$ is a
ramification point if and only if
$k_j>1$. A ramification point is simple if the corresponding ramification index equals $2$.)
The Riemann-Hurwitz formula implies that the dimension of this space is $M=2g+l+N-2$.
One has also the equality $k_1+\dots +k_l=N$.
Two branched coverings $p_1:\L_1\rightarrow {\mathbb P}^1$ and $p_2:\L_2\rightarrow {\mathbb P}^1$
are said to be equivalent if there exists a biholomorphic map $f:\L_1\to \L_2$ such that $p_2f=p_1$.

We also introduce the covering $\hat{H}_{g, N}(k_1, \dots, k_l)$ of the space $H_{g, N}(k_1,
\dots, k_l)$ consisting of pairs
$$<[p:\L\to{\mathbb P}^1]\in H_{g, N}(k_1, \dots, k_l), \{a_\alpha, b_\alpha\}_{\alpha=1}^g>,$$
where $\{a_\alpha, b_\alpha\}_{\alpha=1}^g$ is a canonical basis of cycles on the Riemann surface $\L$.
The spaces $\hat{H}_{g, N}(k_1, \dots, k_l)$ and $H_{g, N}(k_1, \dots, k_l)$ are connected
complex manifolds and the local coordinates on these manifolds are
given by the finite critical values of the map $p$, namely $\l_1, \dots, \l_M$. For $g=0$ the spaces $
\hat{H}_{g, N}(k_1, \dots, k_l)$ and $H_{g, N}(k_1, \dots, k_l)$
coincide.

Let $\phi$ be a primary differential (see \cite{Dubrovin}) on the Riemann surface $\L$.
With this one may induce the structure of a semisimple Frobenius manifold $M_\phi$ on  $\hat{H}_{g, N}(k_1, \dots, k_l)$
by defining: the multiplication law on the tangent bundle $\partial_{\l_m}\circ \partial_{\l_n}=
\delta_{mn}\partial_{\l_m}$; the unity $e=\sum_{m=1}^M\partial_{\l_m}$; the Euler field $
E=\sum_{m=1}^M\l_m\partial_{\l_m}$ and the one-form $\Omega_{\phi^2}=\sum_{m=1}^M
\{{\rm Res}_{P_m}(\phi^2/d\l)\}d\l_m$, where $\l$ is the coordinate on the Riemann surface $\L$
lifted from the base ${\mathbb P}^1$.

The invariant metric $\xi(v, w)=\Omega_{\phi^2}(v\circ w)$ on the Frobenius manifold is flat and
potential. In the coordinates $\l_1, \dots, \l_M$  this metric is
diagonal: $\xi=\sum_{m=1}^M\xi_{mm}(d\l_m)^2$, $\xi_{mm}={\rm Res}_{P_m}(\phi^2/d\l)$.
The rotation coefficients, $\gamma_{mn}$ ($m\neq n$), of this metric are defined by the equality
$$\gamma_{mn}=\frac{\partial_{\l_n}\sqrt{\xi_{mm}}}{\sqrt{\xi_{nn}}}.$$
The Jacobian (\ref{Jacobian}) of the transformation between the flat and canonical
coordinate systems is given by
\[
J^2 = \prod_{m=1}^M \xi_{mm} = \prod_{m=1}^M {\rm Res}_{P_m}(\phi^2/d\l)\,.
\]

\subsubsection{The isomonodromic tau-function of a semisimple Frobenius manifold}

Let $M$ be a semisimple Frobenius manifold with canonical coordinates $\l_1, \dots, \l_M$
and invariant metric $\xi$ having rotation coefficients $\gamma_{mn}$.
Let $\Gamma=||\gamma_{mn}||_{m, n=1, \dots, M; m\neq n}\,,$ ${\mathcal U}={\rm diag}
(\l_1, \dots, \l_M)$ and
$$V=[\Gamma, {\mathcal U}].$$
The isomonodromic tau-function $\tau_I$ of the semisimple Frobenius manifold $M$ is defined
by the system of (compatible) equations
\begin{equation}\label{taudef}
\frac{\partial \log\tau_I}{\partial \l_m}=H_m, \ \ \ m=1, \dots, M,
\end{equation}
where the {\it quadratic Hamiltonians} $H_m$ are defined by
\begin{equation}\label{Hamilt}
H_m=\frac{1}{2}\sum_{n\neq m; 1\leq n\leq M}\frac{V_{nm}^2}{\l_m-\l_n}, \ \ m=1, \dots, M.
\end{equation}

\begin{remark} {\rm In case of the Frobenius manifold $M_\phi$ from section \ref{sec1}
the rotation coefficients $\gamma_{mn}$ and the quadratic Hamiltonians are independent
of the primary differential $\phi$, i.e. are the same for all Frobenius structures on
the Hurwitz space $\hat{H}_{g, N}(k_1, \dots, k_l)$.}
\end{remark}

\begin{remark} {\rm The matrices $A_k$ that appear in isomonodromy problem (\ref{Schlesinger})
are defined in terms of the $V_{mn}$ by
$A_k=||a^k_{mn}||_{m, n=1, \dots M}$ where
$a^k_{mn}=0$ if $m\neq k$ and $a^k_{kn}=V_{kn}$ for $n=1, \dots M$.}
\end{remark}

\begin{remark} {\rm The function $\tau_I$ is only defined up to an overall multiplicative constant. Similarly
the $G$ function is only defined up to an overall additive constant. Such numerical
factors (and a multiplicative factor in $J$) will be ignored.}
\end{remark}

\subsubsection{Quadratic Hamiltonians and Hurwitz spaces}
If a semisimple Frobenius manifold $M$ is constructed by means of a Hurwitz space
then the corresponding quadratic Hamiltonians (\ref{Hamilt}) admit an alternative expression first found
in \cite{KokKor1}.
To give this expression we need to introduce the so-called Bergmann projective connection on the Riemann
surface $\L$ and choose a special system of local parameters on $\L$ connected with
the covering $p:\L\to {\mathbb P}^1$ from the space $H_{g, N}(k_1, \dots, k_l)$.

First recall the definition of the Bergmann kernel.
In the case $g>0$ the Bergmann kernel on the Torelli marked Riemann surface $\L$ is defined
by $B(P, Q)=d_Pd_Q\log E(P, Q),$
where $E(P, Q)$ is the prime-form on $\L$ (see \cite{Fay}).
On the diagonal $P=Q$ the Bergmann kernel is singular:
\begin{equation}\label{funH}
B(x(P), x(Q))=\left(\frac{1}{(x(P)-x(Q))^2}+H(x(P), x(Q)\right)dx(P)\,dx(Q),
\end{equation}
where
\begin{equation}\label{Bcon}
H(x(P), x(Q))=\frac{1}{6}S_B(x(P))+o(1)
\end{equation}
as $P\to Q$. Here $x(P)$ is a local coordinate of a point $P\in\L$, $S_B$ is {\it the Bergmann
projective connection} (see, e.g., \cite{Fay},\cite{Tyurin}).
One must emphasize that the Bergmann projective connection
depends on the choice of local parameter.
If $g=0$ and $z:\L\rightarrow{\mathbb P}^1$ is a biholomorphic map then the Bergmann
kernel is defined by
$$B(z(P), z(Q))=\frac{dz(P)dz(Q)}{(z(P)-z(Q))^2}.$$
(In particular $S_B(z)\equiv 0$ in the local parameter $z$.)

Let $\l$ the standard local parameter on ${\mathbb C}P^1\setminus \{\infty\}$.
This local parameter gives rise to a local parameter $\l(P)$
($P\in \L$, $\l(P)=p(P)$, where $p:\L\to {\mathbb P}^1$ is the covering map)
on the Riemann surface $\L$
which is suitable everywhere except in neighbourhoods of the ramification points $P_1, \dots, P_M$ and at the
infinities $\infty_1, \dots, \infty_l$.
In the neighbourhood of the ramification point $P_m$ one defines the
local parameter $x_m(P)$ by the formula $x_m(P)=\sqrt{\l(P)-\l_m}$ and in the
neighbourhood of the point $\infty_s$ one defines the
local parameter $\zeta_s(P)$ by the formula $\zeta_s(P)=(\l(P))^{-1/k_s}\,.$

The following theorem, proved in \cite{KokKor1}, relates the quadratic Hamiltonians
to the Bergmann projective connection:

\begin{theorem}
Let $M_\phi$ be the Frobenius manifold constructed by means of a Hurwitz space
$H_{g, N}(k_1, \dots, k_l)$ and a primary differential $\phi$.
Let the pair
$$<[p:\L\to{\mathbb P}^1]\in H_{g, N}(k_1, \dots, k_l), \{a_\alpha, b_\alpha\}_{\alpha=1}^g>$$
(here $\{a_\alpha, b_\alpha\}_{\alpha=1}^g$ is a canonical basis of cycles on the Riemann surface $\L$) be a point of $M_\phi$.
The quadratic Hamiltonians $H_m$ are connected with the Bergmann projective connection
on the Riemann surface $\L$ as follows
\begin{equation}\label{osnkk}
H_m=\frac{1}{24}S_B(x_m)\Big|_{x_m=0}; \ \ m=1, \dots, M.
\end{equation}
\end{theorem}

\noindent Thus, the isomonodromic tau-function of the Hurwitz related Frobenius manifold $M_\phi$
is subject to the system of equations
\begin{equation}\label{syst}
\frac{\partial \log\tau_I}{\partial \l_m}=\frac{1}{24}S_B(x_m)\Big|_{x_m=0}; \ \ m=1, \dots, M.
\end{equation}
In the next section we integrate this system explicitly for spaces of branched coverings
of genus zero and one.

\subsection{Calculation of $\tau_I$}
\subsubsection{Rauch variational formula}

Let $p:\L\to {\mathbb P}^1$ be an $N$-fold covering
(or more precisely, a representative of the class of equivalent coverings).
Moving the critical values $\l_1, \dots, \l_M$ of the map $p$, we deform the covering
$\L$. The behaviour of (normalized) holomorphic differentials on $\L$ under such a deformation
is described by the classical Rauch variational formula
(see, e.g., \cite{KokKor}
for the proof of a more general result) which can be written as follows.

Let $g\geq 1$ and let $v$ be a  holomorphic differential on $\L$  with fixed $a$-periods.
(Recall that we are considering  deformations of our covering;
this  means, in particular, that $v$ is {\it a family} of holomorphic differentials on
Torelli marked Riemann surfaces.)  Then
\begin{equation}\label{Ra}
\frac{\partial v}{\partial \l_k}(y)=\frac{1}{2}B(y, x_k)v(x_k)(dx_k)^{-2}|_{x_k=0}.
\end{equation}
In the $g=0$ case $\L$ is a rational surface and there exists a unique biholomorphic map
$U:{\L}\rightarrow {\mathbb P}^1$
such that $U(P)=[\l(P)]^{1/k_1}+o(1)$ as $P\to\infty_1$. (This asymptotic condition replaces
fixing the $a$-periods of the differential in $g\geq 1$ case.)
Then
\begin{equation}\label{Ra0}
\frac{\partial \{dU\}}{\partial\l_k}(y)=\frac{1}{2}B(y, x_k)dU(x_k)(dx_k)^{-2}|_{x_k=0},
\end{equation}
where $B(z_1, z_2)=\frac{dz_1dz_2}{(z_1-z_2)^2}, \ z_1, z_2\in {\mathbb P}^1$
is the Bergmann kernel on the Riemann sphere.
Let
\begin{eqnarray*}
b(P_k, P_l) & = & [B(x_l, x_k)dx_l^{-1}dx_k^{-1}]\Big|_{x_k=0, x_l=0}\,,\quad\quad\ \ \ k\neq l,\\
b(P_l, \infty_s) & = &[B(x_l, \zeta_s)dx_l^{-1}d\zeta_s^{-1}]\Big|_{x_l=0, \zeta_s=0}.
\end{eqnarray*}

\noindent The following lemma relates these functions to various derivatives. Ultimately these will be
used to integrate the system (\ref{syst}).

\begin{lemma}\label{not}
Let $dz=f(x_k)dx_k$ be a normalized holomorphic differential on the elliptic surface $\L$ written in the local
parameter near the point $P_k$ and let
$$f(x_k)=f_k+f_{k, 1}x_k+f_{k,2}x_k^2+O(x_k^3)$$
as $x_k\to 0$.
The function $h(\zeta_s)$ is defined in a neighborhood of the point $\infty_s$
by $dz=h(\zeta_s)d\zeta_s$ and let $h_s=h(\zeta_s)|_{\zeta_s=0}$.

\medskip

\noindent With these definitions

\begin{eqnarray}
\frac{\partial f_l}{\partial \l_k} & =& \frac{1}{2}b(P_k, P_l)f_k\,,\quad\quad \ \
k\neq l,
\label{eq}\\
\frac{\partial h_s}{\partial \l_k} & =& \frac{1}{2}b(P_k, \infty_s)f_k,
\label{eqq}
\end{eqnarray}

\noindent and
\begin{equation}\label{eq1}
\frac{\partial\log f_k}{\partial \l_k}=\frac{1}{12}S_B(x_k)|_{x_k=0}+\frac{f_{k, 2}}{2f_k}.
\end{equation}
In the rational case
equations (\ref{eq}), (\ref{eqq})
and (\ref{eq1}) hold with the function $f(x_k)$ defined by $dU(x_k)=f(x_k)dx_k$
and the function $h(\zeta_s)$ defined by $dU(\zeta_s)=h(\zeta_s)d\zeta_s$.
If $g=0$ the number $h_s$ is defined only for $s>1$.

\end{lemma}
{\bf Proof.} Equations (\ref{eq}) and (\ref{eqq}) immediately follow from the Rauch formula.
To prove (\ref{eq1}) observe that
$$f(x_k)dx_k=\left(f_k+f_{k, 1}x_k+f_{k, 2}x_k^2+O(x_k^3)\right)\frac{d\l}{2\sqrt{\l-\l_k}}$$
and
\begin{equation}\label{pro}
\frac{\partial\{f(x_k)dx_k\}}{\partial \l_k}=\left[\frac{\partial f_k}{\partial \l_k}
-\frac{f_{k, 2}}{2}\right]+
\frac{f_k}{2x_k^2}+o(1)
\end{equation}
 as $x_k\to 0$.
On the other hand, owing to the Rauch formula and the asymtotics of the Bergmann kernel at the diagonal,
one has
\begin{equation}\label{pro1}
\frac{\partial\{f(x_k)dx_k\}}{\partial \l_k}=\frac{f_k}{2x_k^2}+\frac{f_k}{12}S_B(x_k)+o(1)
\end{equation}
as $x_k\to 0$.
Comparing (\ref{pro}) and (\ref{pro1}), one obtains (\ref{eq1}).$\square$

\subsubsection{The Main Theorem}

Using the notation of Lemma \ref{not} one may now write down expressions for the isomonodromic tau-function
for low genus Hurwitz spaces.

\begin{theorem}\label{maint}
The isomonodromic tau-function corresponding to the Frobenius structures on the space $H_{0, N}(k_1, \dots, k_l)$
is given by the expression
\begin{equation}\label{eqq0}
\tau_I=\left\{\frac{\prod_{m=1}^Mf_m}{\prod_{s=2}^lh_s^{k_s+1}}\right\}^{\frac{1}{24}}.
\end{equation}
The isomonodromic tau-function corresponding to the Frobenius structures on the space $H_{1, N}(k_1, \dots, k_l)$
is given by the expression
\begin{equation}\label{eqq1}
\tau_I=\frac{1}{\eta(\sigma)}\left\{\frac{\prod_{m=1}^Mf_m}{\prod_{s=1}^lh_s^{k_s+1}}\right\}^{
\frac{1}{24}},
\end{equation}
where $\sigma$ is the modulus of the elliptic surface $\L$, $\eta$ is the Dedekind eta-function.
\end{theorem}
\subsubsection{Proof of the main theorem}

We recall some useful relations from \cite{Fay}. (In agreement with the paper \cite{Fay2},
we use here the normalization condition $(\int_av, \int_bv)=(1, \sigma)$ for the normalized
holomorphic differential $v$ on the elliptic surface $\L$. In \cite{Fay} the
normalization $(\int_av, \int_bv)=(2\pi i, \sigma)$ was used and this explains why the
formulae here differ slightly from those in \cite{Fay}.)

Let $\L$ be a marked elliptic surface and let the $b$-period of the normalized differential
$v$ be $\sigma$.
Introduce the function $\E$ by the equation
$$\E(\sigma)=\frac{d}{d\sigma}\log{\eta(\sigma)},$$
where $\eta$ is the Dedekind eta-function.
Owing to the heat equation for theta-functions we have
\begin{equation}\label{E}
\E(\sigma)=\frac{1}{12\pi i}\frac{\theta'''\left[^{\frac{1}{2}}_{\frac{1}{2}}\right](0|\sigma)}
{\theta'\left[^{\frac{1}{2}}_{\frac{1}{2}}\right](0|\sigma)}.
\end{equation}
The Bergmann kernel of the elliptic surface $\L$ is given by
\begin{equation}\label{ellB}
B(x, y)=\left[\P(\int_x^yv)-4\pi i\E(\sigma)\right]v(x)v(y),
\end{equation}
where $\P$ is the Weierstrass $\P$-function.
Let $z$ be the coordinate on the universal covering ${\mathbb C}$ of the elliptic surface $\L$.
For the invariant Wirtinger projective connection $S_W$  (see \cite{Fay}) we have
\begin{equation}\label{Wirt}
S_W(x)=S_B(x)+24\pi i\E v^2(x)=\{z, x\},
\end{equation}
where $\{z, x\}$ is the Schwarzian derivative (see also \cite{Tyurin}, the last example in \S1.3).
Note that for a rational surface $\L$ the Bergmann projective connection coincides with the Wirtinger invariant
projective connection and is nothing but the Schwarzian derivative of the map $U$:
\begin{equation}\label{BergSph}
S_W(x)=S_B(x)=\{U(x), x\}.
\end{equation}

\begin{proposition}\label{mainth}
\[\phantom{tqftftf}\]
\begin{itemize}
\item Let $g=0$.
Define the function ${\bf T_0}(\l_1, \dots, \l_M)$ by
\begin{equation}\label{T0}
 {\bf T_0}(\l_1, \dots, \l_M)=\log\left\{\frac{\prod_{m=1}^Mf_m}{\prod_{s=2}^lh_s^{k_s+1}}\right\}.
\end{equation}
Then for any $k=1, \dots, M$
\begin{equation}\label{T01}
\frac{\partial {\bf T_0}}{\partial \l_k}=S_W(x_k)|_{x_k=0}=S_B(x_k)|_{x_k=0}=\{U(x_k), x_k\}|_{x_k=0}.
\end{equation}
\item Let $g=1$.
Define the function ${\bf T_1}(\l_1, \dots, \l_M)$ by
\begin{equation}\label{T1}
 {\bf T_1}(\l_1, \dots, \l_M)=\log\left\{\frac{\prod_{m=1}^Mf_m}{\prod_{s=1}^lh_s^{k_s+1}}\right\}.
\end{equation}
Then for any $k=1, \dots, M$
\begin{equation}\label{T11}
\frac{\partial {\bf T_1}}{\partial \l_k}=S_W(x_k)|_{x_k=0}=S_B(x_k)|_{x_k=0}+24\pi i\E f_k^2=\{z, x_k\}|_{x_k=0}.
\end{equation}
\end{itemize}

\end{proposition}

\noindent {\bf Proof.} Using Lemma \ref{not} and formula (\ref{ellB}) one obtains
\begin{eqnarray*}
\partial_{\l_k}{\bf T_1} & = &\frac{1}{2}\sum_{m\neq k}\frac{b(P_m, P_k)f_k}{f_m}+\frac{1}{12}S_B(x_k)\Big|_{x_k=0}+
\frac{f_{k, 2}}{2f_k}-\frac{1}{2}\sum_{s=1}^l(k_s+1)\frac{b(P_k, \infty_s)f_k}{h_s}\,,\\
& = &\left\{\frac{1}{2}\sum_{m\neq k}\P(\int_{P_m}^{P_k}v)-
\frac{1}{2}\sum_{s=1}^l(k_s+1)\P(\int_{\infty_s}^{P_k}v)\right\}f_k^2+
\frac{1}{12}S_B(x_k)\Big|_{x_k=0}\\
&  &+\frac{f_{k, 2}}{2f_k}-2\pi i(M-1)\E f_k^2+
2\pi i\sum_{s=1}^l(k_s+1)\E f_k^2.
\end{eqnarray*}
Using (\ref{Wirt}), together with the Riemann-Hurwitz formula $M=l+\sum_{s=1}^lk_s$, this simplifies to
\begin{equation}\label{ee2}
\partial_{\l_k}{\bf T_1}=\frac{1}{2}\left\{\sum_{m\neq k}\P(\int_{P_m}^{P_k}v)-
\sum_{s=1}^l(k_s+1)\P(\int_{\infty_s}^{P_k}v)\right\}f_k^2+\frac{f_{k, 2}}{2f_k}+\frac{1}{12}S_W(x_k)|_{x_k=0}.
\end{equation}
The analogous formula in the $g=0$ case is
\begin{equation}\label{ff1}
\partial_{\l_k}{\bf T_0}=
\frac{1}{2}\left\{\sum_{m\neq k}\frac{1}{(z_k-z_m)^2}-\sum_{s=2}^l\frac{k_s+1}{(z_k-y_s)^2}\right\}f_k^2+
\frac{f_{k, 2}}{2f_k}+\frac{1}{12}S_W(x_k)|_{x_k=0}.
\end{equation}
where $z_m=U(P_m); m=1, \dots, M$ and $y_s=U(\infty_s); s=2, \dots, l$.

In the $g=0$ case let $R:{\mathbb P}^1\rightarrow {\mathbb P}^1$ be the composition $p\circ U^{-1}$
(we recall that
$p:\L\to {\mathbb P}^1$ is the chosen covering from the Hurwitz space).
Then $R$ is a rational function and the expression in large braces in (\ref{ff1}) coincides with
\begin{equation}\label{d1}
\left[-\frac{d}{dz}\left(\frac{R''(z)}{R'(z)}\right)-\frac{1}{(z-z_k)^2}\right]\Bigg|_{z=z_k},
\end{equation}
Similarly, in the $g=1$ case let $\Delta$ be a fundamental parallelogram on the universal covering ${\mathbb C}$ of
the elliptic surface $\L$ and let ${\mathcal U}^{-1}:\Delta\rightarrow \L$ be the uniformization map.
Let ${\mathcal R}=p\circ{\mathcal U}^{-1}$. Then ${\mathcal R}$ is an elliptic function and
the expression in large braces in (\ref{ff1}) coincides with
\begin{equation}\label{d2}
\left[-\frac{d}{dz}\left(\frac{{\mathcal R}''(z)}{{\mathcal R}'(z)}\right)-\P(\int_{z_k}^z\,dz)\right]\Bigg|_{z=z_k}.
\end{equation}
Here we are denoting the standard local parameters on ${\mathbb P}^1$ and $\Delta$ by $z$
and, in the last equation, $z_k\in \Delta, \ {\mathcal U}^{-1}(z_k)=P_k$.

From now on the proofs of (\ref{T01}) and (\ref{T11}) coincide verbatim and only the details of the
rational case will be presented. Let $x=z-z_k$ and define $\alpha\,,\beta$ and $\gamma$ by the
condition that
\begin{equation}\label{RR}
R'(z)=\alpha x+\beta x^2+\gamma x^3+O(x^4)
\end{equation}
as $z\to z_k$.
It then follows that
\begin{equation}\label{w1}
\left(\frac{dz}{dx_k}\right)^2=\frac{4(R(z)-R(z_k))}{[R'(z)]^2}=\frac{2}{\alpha}-\frac{8}{3}\frac{\beta}
{\alpha^2}x+\left(\frac{10}{3}\frac{\beta^2}{\alpha^3}-3\frac{\gamma}{\alpha^2}\right)x^2+O(x^3)
\end{equation}
and, in particular, that
\begin{equation}\label{RR1}
f_k^2=\frac{2}{\alpha}.
\end{equation}
Similarly,
\begin{equation}\label{RR4}
\frac{f_{k,2}}{2f_k}=\lim_{x_k\to 0}\frac{z''(x_k)-z''(0)}{4(z(x_k)-z(0))}=\frac{5}{6}\frac{\beta^2}{\alpha^3}
-\frac{3}{4}\frac{\gamma}{\alpha^2}.
\end{equation}
Using the expansion (\ref{RR}) it is straightforward to show that
\begin{eqnarray}\label{RR3}
\frac{2\gamma\alpha-\beta^2}{\alpha^2} & = &
\left[\frac{d}{dz}\left(\frac{R''(z)}{R'(z)}\right)+\frac{1}{(z-z_k)^2}\right]\Bigg|_{z=z_k}\,,\\
& = & \left[\{R(z), z\}+\frac{1}{2}\left(\frac{R''(z)}{R'(z)}\right)^2+
\frac{1}{(z-z_k)^2}
\right]\Bigg|_{z=z_k}\,.
\end{eqnarray}
On using the transformation property of the Schwarzian derivative, together with equations
(\ref{RR1}) and (\ref{w1}), one obtains
\begin{eqnarray*}
\frac{4\gamma\alpha-2\beta^2}{\alpha^3} & = &
\left[
\{R(z),z\}\left(\frac{dz}{dx_k}\right)^2+\left(
\frac{1}{2}\left(\frac{R''(z)}{R'(z)}\right)^2+\frac{1}{(z-z_k)^2}\right)
\left(\frac{dz}{dx_k}\right)^2
\right]\Bigg|_{z=z_k}\,, \\
& = &
\left[
\{R(z), x_k\}-\{z, x_k\}+
\left(
\frac{1}{2}\left(\frac{R''(z)}{R'(z)}\right)^2+\frac{1}{(z-z_k)^2}\right)
\left(\frac{dz}{dx_k}\right)^2
\right]\Bigg|_{z=z_k}\,,\\
& = & \frac{\gamma}{\alpha^2}-\{z, x_k\}\Big|_{x_k=0}\,,
\end{eqnarray*}
so
\begin{equation}\label{SS}
\{z, x_k\}\Big|_{x_k=0}=\frac{2\beta^2-3\alpha\gamma}{\alpha^3}.
\end{equation}
Finally, substituting into (\ref{ff1}) the expressions
(\ref{RR3}), (\ref{RR4}) and (\ref{SS}), we get
$$\partial_{\l_k}{\bf T_0}=
\frac{1}{12}\left(\frac{2\beta^2-3\alpha\gamma}{\alpha^3}\right)-\frac{3}{4}\frac{\alpha\gamma}{\alpha^3}+
\frac{5}{6}\frac{\beta^2}{\alpha^3}+\frac{\beta^2-2\gamma\alpha}{\alpha^3}=
\frac{2\beta^2-3\alpha\gamma}{\alpha^3}=\{z, x_k\}\Big|_{x_k=0}.$$
$\square$

\medskip

\noindent Theorem \ref{maint} immediately follows from equations (\ref{syst}), proposition \ref{mainth}
and the Rauch formula
$$\frac{\partial \sigma}{\partial \l_k}=\pi i f_k^2\,,$$
a proof of which can be found in \cite{KokKor}.

\section{The space of branched coverings of genus 0}

As commented on earlier, Hurwitz spaces $H_{0,N}(k_1\,\ldots,k_l)$ are particularly
simple: firstly the spaces $\hat{H}_{g, N}$ and $H_{g, N}$
coincide and secondly the maps $p:\mathbb{P}\rightarrow\mathbb{P}$ are just rational
functions. Such a rational map may, in a suitable coordinate system, be taken to
be\footnote{In this formula for $p\,,$ and in
various formulae to come, slightly different forms are required if
$k_1=1\,,$ or in some, if $k_i=1\,.$ The final results, though,
are not sensitive to this, so only the generic calculations will
be presented here.}
\[
p(z)= z^{k_1} + \sum_{r=0}^{k_1-2} a_r z^r - \sum_{i=2}^l
\sum_{\alpha^i=1}^{k_i}
\frac{c_{(i,\alpha^i)}}{(z-b^i)^{\alpha^i}}\,,
\]
and in order for $p$ to be well defined the poles $b^i$ must be
distinct and the coefficients $c_{(i,k_i)}$at the end of the various Laurent
tails must be non-zero. Thus
\[
H_{0,N}(k_1\,,\ldots\,,k_l)\cong\mathbb{C}^M \backslash \{ b^i-b^j=0\,, i
\neq j\} \cup \{ c_{(i,k_i)} = 0 \}\,.
\]
The main additional object is a choice of a primitive form. Here
this may be taken to be $\phi=dz\,.$ With this the flat
coordinates
\[
\{t^{(i,\alpha^i)}\,:i=1\,,\ldots,l\,,\alpha^i=2\,,\ldots\,,k_i\}
\cup \{p^i\,,q^i\,:i=2\,,\ldots\,,l\}
\]
are given by Theorem 5.1 of \cite{Dubrovin}. For future notational
convenience we define $t_i=t^{(i,k_i)}\,.$ With the above formulae
for $p$ and $\phi$ one may evaluate these formulae,
obtaining, in particular
\[
\left.
\begin{array}{rcl}
p^i & = & b^i \,,\\
t_i & = & k_i \, {c_{(i,k_i)}}^{{1/k_i}}
\end{array}
\right\} \quad i=2\,,\ldots\,,l\,,
\]
so
\[
H_{0,N}(k_1\,,\ldots\,,k_l)\cong\mathbb{C}^M  \backslash \mathcal{S}_1\cup\mathcal{S}_2
\]
where
\begin{eqnarray*}
\mathcal{S}_1 & \cong & \{ p^i-p^j=0\,, i\neq j\}\,,\\
\mathcal{S}_2 & \cong & \{ t_i = 0\,, i=2\,\ldots\,, l \}\,.
\end{eqnarray*}

\subsection{The $G$-function in flat coordinates}

Theorem 2 together with (\ref{Gfunction}) enables the $G$-function to
be calculated explicitly in flat-coordinates. Owing to some cancellations
it turns out to be easier to calculate the $G$-function than the $\tau_I$-function.
A separate calculation of
the Jacobian $J$ then enables the $\tau_I$-function to be found.

\begin{theorem}

The $G$-function for the Hurwitz space
$H_{0,N}(k_1\,,\ldots\,,k_l)$ is given by
\begin{equation}
G=-\frac{1}{24} \sum_{i=2}^l (k_i+1) \log t_i\,. \label{GHurwitz0}
\end{equation}
The scaling anomaly $\gamma=\mathcal{L}_EG$ is given by
\[
\gamma=-\frac{1}{24} \left( l-2 + \sum_{i=1}^l {k_i}^{-1} +
{k_1}^{-1} M\right)\,.
\]
\end{theorem}

\noindent{\bf Proof~} The proof is really only an extension of the
examples in \cite{KokKor1} coupled with the flat-coordinate calculation
given above. The formula for the $G$-function is:
\[
G=\frac{1}{24} \log \left\{ \frac{\prod_{r=1}^{M}
\left.\frac{dz}{dx_r}\right|_{x_r=0}} {\prod_{s=2}^l \left(
\frac{dz}{d\xi_s}\right)^{k_s+1} \left( \prod_{r=1}^{M}
\res_{P_r}
\frac{\phi^2}{d\lambda}\right)^{\frac{1}{2}}}\right\}\,.
\]
Since
\[
p=\frac{c_{(i,k_i)} + O(z-b^i)}{(z-b^i)^{k_i}}
\]
and $p=\xi^{-k_i}$ around $b^i$ one may obtain
\[
\left.\frac{dz}{d\xi_i}\right|_{\xi_i=0} =
c_{(i,k_i)}^{k_i^{-1}}\,.
\]
Also
\[
\res_{P_m} \frac{\phi^2}{d\lambda} = \frac{1}{2^M} \,
\left.\frac{dz}{dx_m}\right|_{{x_m=0}}\,,
\]
so the terms which are hard to compute in the numerator and
denominator cancel, leaving only a products which have already
been evaluated. Hence
\[
G=-\frac{1}{24} \sum_{m=2}^l \left( \frac{k_i+1}{k_i} \right)
\log c_{(i,k_i)}\,.
\]
Expressing this in flat coordinates yields the final result. Note that $G$ is non-singular at all points in
$H_{0,N}(k_1\,\ldots\,k_l)\,;$ the singularities occurring only on the
component $\mathcal{S}_2$ of the boundary of the Hurwitz space.
$\square$

\medskip
\bigskip

\bigskip

In the special case $H_{0,N}(N)\,,$ which is isomorphic to the orbit
space $\mathbb{C}^{N-1}/A_{N-1}\,,$ one obtains $G=0$ and for the
case $H_{0,N}(k,N-k)$ which is isomorphic to the orbit
space
$\mathbb{C}^{N}/{\widetilde{W}}^{(k)}{(A_{N-1})}\,,$
where ${\widetilde{W}}^{(k)}{(A_{N-1})}$ is an extended affine
Weyl group \cite{DZ2}, one obtain the formulae derived in \cite{Strachan},
though in a slightly different form due to a different primitive form having being used there.
Such a change in the primitive form induces a symmetry in the
corresponding Frobenius manifolds and this in turn induces a
simple transformation for the $G$-function (see Lemma 9 in \cite{Strachan}).

\subsection{The $\tau_I$-function in flat coordinates}

For $p(z)$ given above, its derivative may be written
uniquely as
\[
p^\prime(z) = \frac{f(z)}{g(z)}
\]
where
\[
g(z) = \prod_{i=2}^l (z-b^i)^{k_i+1}\,.
\]
The canonical coordinates are given in terms of the zeroes of
$p^{\prime}\,,$
\[
\left.
\begin{array}{rcl}
p^{\prime}(\alpha^i) & = & 0 \\
p(\alpha^i)& = & \lambda^i
\end{array}
\right\} \qquad\quad i=1\,,\ldots M\,.
\]
By calculating the resultant of $f$ and $g$ one can see that they
have no common roots on $H_{0,N}(k_1\,,\ldots\,,k_l) $ and so the zeroes of
$p{^\prime}$ are precisely the roots of the polynomial
$f\,.$

\begin{lemma}
\[
{\mathcal{R}}(f,g) = \left\{ \prod_{i\neq j}
(p^i-p^j)^{(k_i+1)(k_j+1)} \right\}\, \left\{ \prod_{i=1}^m
t_i^{k_i(k_i+1)} \right\}\,.
\]
\end{lemma}

\noindent{\bf Proof} Writing $f$ as
\[
f(z) = k_1 \prod_{i=1}^{M} (z-\alpha^i)
\]
gives
\[
{\mathcal{R}}(f,g) = \prod_{i=1}^{M} \prod_{j=2}^l
(\alpha^i-p^i)^{k_i+1}\,.
\]
But from differentiating the expansion of $p\,,$
\[
c_{(i,k_i)} = \frac{ \prod_{j=1}^{M}
(p^i-\alpha^j)}{\prod_{i\neq j} (p^i-p^j)^{k_j+1}}
\]
Using the expression of $c_{(i,k_i)}$ in terms of flat coordinates
completes the proof.$\square$

\medskip

\bigskip

\bigskip

\noindent Note that ${\mathcal{R}}(f,g)$ is a non-zero function on
$H_{0,N}(k_1\,\ldots\,k_l)\,:$ it vanishes precisely on boundary $\mathcal{S}_1\cup\mathcal{S}_2\,.$ The
values of $p^{\prime\prime}(\alpha^i)\,,$ or rather the
product of these values, are now easily calculated using basic
properties of the resultant:
\[
p^{\prime\prime}(\alpha^i) =
\frac{f^{\prime}(\alpha^i)}{g(\alpha^i)}\,,
\]
and hence
\[
\prod_{i=1}^{M} p^{\prime\prime}(\alpha^i) =
\frac{\prod_{i=1}^{M}
f^{\prime}(\alpha^i)}{\prod_{i=1}^{M} g(\alpha^i)}=
\frac{\mathcal{R}(f,f^{\prime})}{\mathcal{R}(f,g)}\,.
\]
This then gives the Jacobian $J\,.$ Finally, putting equation
(\ref{Gfunction}), Theorem 2 and Lemma 3 together gives:

\begin{theorem} The isomonodromic $\tau$-function for the space
$H_{0,N}(k_1\,,\ldots\,,k_l)$ is given by the formula
\[
\tau_I^{-48} = \frac{ \mathcal{R}(f,f^{\prime}) } {\left\{
\prod_{i\neq j} (p^i-p^j)^{(k_i+1)(k_j+1)} \right\}\, \left\{
\prod_{i=2}^l t_i^{(k_i+1)(k_i-2)} \right\}}\,.
\]
\end{theorem}
\noindent This formula shows a number of different things.
Firstly, $\tau_I^{-48}$ is the ratio of quasi-homogeneous
polynomials. Moreover the denominator is non-zero at all points in
$H_{0,N}(k_1\,\ldots\,k_l)\,,$ vanishing only on the boundary $\mathcal{S}_1\cup\mathcal{S}_2\,.$ Hence the
zero locus
\[
\{ {\bf t}\in\M\,: \tau_I^{-48} = 0 \}  = \{ \mathcal{R}(f,f')=0
\} = \K
\]
that is, the classical caustic where two or more of the $\alpha^i$
are coincide. Thus $\tau_I^{-48}$ is non-zero on $\M\backslash\K\,$ and
zero only on $\K\,.$ Such a caustic is stratified by the
multiplicities of the roots of $f(z)=0\,.$

\medskip

\noindent The quasihomogeneous polynomial $\mathcal{R}(f,f')$ is
not irreducible:

\begin{proposition}
\[
\mathcal{R}(f,f')= \left\{\prod_{\{i\,:k_i=1\}} t_i \right\}\cdot
\left\{\prod_{r < s} (p^r-p^s)^{m_{rs}} \right\}\cdot
{\widetilde{\mathcal{R}}}(f,f')\,.
\]
\end{proposition}

\noindent{\bf Proof~} To derive this we use the property of the
resultant that there exists functions $\alpha(z)$ and $\beta(z)$
(which here will be quasihomogeneous polynomials) such that
\[
\mathcal{R}(f,f') = \alpha(z) f(z) + \beta(z) f'(z)\,.
\]
The function $f$ is explicitly
\[
f(z) = \prod_{i=2}^l (z-b^i)^{k_i+1} \cdot \left\{ k_1 z^{k_1-1}
+ \ldots + \sum_{i=2}^l \sum_{\alpha^i=1}^{k_i} \frac{\alpha^i
c_{(i,\alpha^i)}}{(z-b^i)^{\alpha^i+1}}\, \right\}
\]
and this may be used to calculate $f(b^i)$ and $f'(b^i)\,.$ For
example
\[
f(b^i) = t_i \prod_{r\neq i} (b^i-b^r)^{k_r+1}
\]
and a similar calculation show that $f'(b^i)$ has a factor $t_i$
if and only if $k_i=1\,.$ Hence $\mathcal{R}$ has a factor of
$t_i$ if and only if $k_i=1\,.$ A similar argument gives the
second factor in the above formula with $m_{rs}=(k_r+k_s)\,.$
$\square$

\medskip
\bigskip

\bigskip

\noindent The $m_{rs}$ factor is not optimal, as may be shown by
calculating certain examples. It seem likely, based on the
explicit calculation of low dimensional examples, that with an
optimal factor, the residual polynomial
${\widetilde{\mathcal{R}}}(f,f')$ is irreducible.

\bigskip

\section{The space of branched coverings of genus 1}

Elliptic functions may easily be manipulated in the same way as
rational functions. Here the definitions and properties
of these and related functions will follow \cite{WW}. Any elliptic
function may be written as
\[
p(z) = a + \sum_{i=1}^l \sum_{\alpha^i=1}^{k_i} c_{(i,\alpha_i)} \zeta^{(\alpha^i-1)}(z-b_i)
\]
with the local expansion around the pole $z=b_i\,,$
\[
\frac{c_{i,1}}{z-b_i} - \frac{c_{(i,2)}}{(z-b_i)^2} + \ldots +
\frac{(k_i-1)!(-1)^{k_i-1} c_{(i,k_i)}}{(z-b_i)^{k_i}}\,.
\]
In order for this to be well-defined, the poles must be distinct with well-defined
orders, so singular spaces $\mathcal{S}_1$ and $\mathcal{S}_2$ may be
defined as above. Such an elliptic function admits an alternative
representation in terms of products of $\sigma$-functions,
\[
p(z)=a_0 \frac{ \prod_{i=1}^N \sigma(z-a_i)}{ \prod_{j=1}^l \sigma(z-b_j)^{k_j}}\,
\]
with $\sum_{i=1}^N a_i = \sum_{j=1}^l k_j b_j\,.$
To proceed further one requires an elliptic version of the resultant of
two polynomials. This may be defined as follows. Let
\[
F(z)=a_0 \prod_{i=1}^M \sigma(z-a_i) \,, \quad\quad
G(z)=b_0 \prod_{j=1}^N \sigma(z-b_j) \,,
\]
and define $\mathcal{R}(F,G)  =  a_0^N \prod_{i=1}^M G(a_i)\,.$ It follows immediately
from this that
\[
\mathcal{R}(F,G)  =  a_0^N b_0^M \prod_{i,j} \sigma(a_i-b_j)
 = (-1)^{MN} \prod_{j=1}^N F(b_j)
 =  (-1)^{MN} \mathcal{R}(G,F)\,.
\]
In particular, $\mathcal{R}(F,G)=0$ if and only if $F$ and $G$ have,
modulo the lattice, a common zero, this following from properties of
the $\sigma$-function. For arbitrary $F$ and $G$ of this form, the
elliptic resultant $\mathcal{R}(F,G)$ does not transform invariantly
under the lattice transformations $a_i\mapsto a_i+ 2 m \omega_1+ 2 n \omega_2\,,$
$b_i\mapsto b_i+ 2 m \omega_1+ 2 n \omega_2\,.$ However, if $F/G$ is
an elliptic function then the elliptic resultant transforms as
\[
\mathcal{R}(F,G) \mapsto c_L(m,n)\, \mathcal{R}(F,G)
\]
where $c_L(m,n)$ is a non-zero function, depending on the periods.

\medskip

The $G$-function and $\tau_I$-function for the space $H_{1,N}(k_1,\ldots\,,k_l)$
may now be calculated in manner entirely analogous to Theorem 3 and 4.

\begin{theorem}

The $G$-function for the Hurwitz space
$H_{1,N}(k_1\,,\ldots\,,k_l)$ is given by
\begin{equation}
G=-\log\eta(t_0)-\frac{1}{24} \sum_{i=2}^l (k_i+1) \log t_i\,.
\label{GHurwitzone}
\end{equation}
The scaling anomaly $\gamma=\mathcal{L}_EG$ is given by
\[
\gamma=-\frac{1}{24} \left( l+ \sum_{i=1}^l {k_i}^{-1} +
{k_1}^{-1} M\right)\,.
\]
\end{theorem}

\noindent{\bf Proof~} The proof is, again, really only an extension of the
example in \cite{KokKor1}, so
\[
\left( \prod_{m=1}^M {\rm Res}_{P_m} \frac{\phi^2}{d\lambda}\right) = {\rm const~}
\left( \prod_{m=1}^M \frac{\omega(x_m(P))}{dx_m(P)}\Bigg|_{P=P_m}\right)^2=\left(\prod_{m=1}^M f_m\right)^2\,,
\]
and the flat coordinate calculations
\[
t_0 = \oint_b \omega = \sigma
\]
and
\begin{eqnarray*}
t_i  & = & t^{(i,\alpha_i)}= {\rm Res}_{z=0} (z[\lambda(z)]^{-\frac{k_i-1}{k_i}} d\lambda(z)) =
{\rm Res}_{\zeta_i=0} \left( z(\zeta_i) \frac{d\zeta_i}{\zeta_i^2} \right)=z'(\zeta_i)|_{\zeta_i=0}=h_i\,,\\
& = & {\rm const~} {c_{(i,k_i)}}^{1/k_i}\,.
\end{eqnarray*}
$\square$

\medskip

\medskip

\noindent In particular, for the space $H_{1,N}(N)$ one obtain the $G$-function conjectured in \cite{Strachan}
and proved in \cite{KokKor1} (N.B. a $log$-term went missing in \cite{KokKor1} and a different
normalization was used \cite{Strachan} which accounts for the different numerical factor in front of the
first term in the $G$-function above).

\medskip

Since $p'(z)$ is also an elliptic function it may be written as $p'=f/g$ where $f$ and $g$
are products of $\sigma$-functions. The proof of Theorem 4 goes through verbatim with
the elliptic resultant introduced above. In particular, the function $\mathcal{R}(f,g)$ can
be calculated in exactly the same way as above.

\begin{theorem} The isomonodromic $\tau$-function for the space
$H_{1,N}(k_1\,,\ldots\,,k_l)$ is given by the formula
\[
\tau_I^{-48} = \frac{  \eta(t_0)^{48} \, \kappa } {\left\{
\prod_{i\neq j} (b^i-b^j)^{(k_i+1)(k_j+1)} \right\}\, \left\{
\prod_{i=1}^l t_i^{(k_i+1)(k_i-2)} \right\}}\,
\]
where
\[
\kappa=
\prod_{r\neq s} \sigma(\alpha_r-\alpha_s)\,
\]
and $\alpha_i\,,\,i=1\,,\ldots\,,M\,,$ are the critical points of the map $p\,.$
\end{theorem}

\medskip

\noindent The submanifold $\kappa^{-1}(0)$ corresponds to the caustic $\mathcal{K}$ where the
multiplication is non-semisimple.

\medskip

\noindent{\bf Example} The Hurwitz space $H_{1,2}(2)$ was studied in detail in \cite{Dubrovin}, appendix C.
Detailed calculations give
\[
\tau_I^{-48} = t_1^{12} \hat\Delta^3(t_0)
\]
where $\hat\Delta(\omega'/\omega)=\omega^{12} \Delta(\omega,\omega')$ and $\Delta$ is the classical
discriminant of the underlying elliptic curve with periods $\omega$ and $\omega'\,.$
This shows that $\tau_I$ is singular where $\Delta=0\,,$
that is, where the underlying elliptic curves develops a singularity. Such points are not strictly
speaking in the Hurwitz space - they occur on its boundary.

\section{Comments}

While for any individual Frobenius manifold one may find the corresponding $G$-function by directly
solving the defining differential equation, calculating it explicitly for classes of
Frobenius manifolds is more problematic. For orbit spaces $\mathbb{C}^M/W$ where $W$ is
a Coxeter group, the $G$-function was calculated by studying the $F$-manifold structure
on the caustics. Such an approach could be applied to other classes of manifolds, such
as the Hurwitz spaces in this paper, but this is hampered by the lack of information on
the $F$-manifold structure on the caustics. The formulae (\ref{GHurwitz0},\ref{GHurwitzone})
suggest that one has a number of logarithmic caustics $\K_{\rm log}$ and that the multiplication
on them is of the type given in Lemma 2.3 in \cite{Strachan} with $N_{\rm log}=(k_i+1)\,,$
but to verify this would require much detailed calculation.
The methods used here are very specific to Hurwitz spaces. However certain
structures appear in both approaches, namely the role of caustics and
singular structures (which include the logarithmic caustics introduced in
\cite{Strachan}) on the boundary of the Hurwitz spaces. Certainly the results
suggest that a detailed study of the caustics and boundaries of Hurwitz spaces should
be undertaken.

More generally, both constructions have a considerable global character: the results
are all obtained in terms of critical data. This suggests that there might be formulae
for the $G$ and $\tau_I$ functions in terms of the Saito formalism of Frobenius manifolds.

\medskip
\noindent{\bf Acknowledgements}
\medskip

We would like to thank S. Natanzon and C. Hertling and most particularly
D. Korotkin for their various comments and criticisms of this work, and the
Max Planck Institute, Bonn, for its hospitality. IABS would like to thank the
EPSRC (grant GR/R05093) for financial support.

\end{document}